\documentstyle[12pt]{article}
\newcommand{\newc}{\newcommand} 
\newc{\ra}{\rightarrow} 
\newc{\lra}{\leftrightarrow} 
\newc{\beq}{\begin{equation}} 
\newc{\eeq}{\end{equation}} 
\newc{\barr}{\begin{eqnarray}} 
\newc{\earr}{\end{eqnarray}} 

 1

\begin{document} 
\begin{titlepage}

\begin{center}
{\large \bf Searching for Cold Dark Matter. A case of coexistence of 
Supersymmetry and Nuclear Physics.}\\

\vspace{15mm}

J.D. VERGADOS \footnote{ Presented by J.D. Vergados.} and T.S. KOSMAS 

\vspace{7mm}

Theoretical Physics Section, University of Ioannina, GR-45110, Greece.

\end{center}

\vspace{12mm}
	
\begin{abstract}
The direct detection rate for supersymmetric cold dark matter (CDM)
particles is calculated for a number of suitable nuclear targets.
Both the coherent and spin contributions are considered. By considering 
representative phenomenologically acceptable input in the restricted 
SUSY parameter space, detectable rates are predicted for some choices 
of the parameters. The modulation effect, due to the Earth's annual motion, 
has also been considered and found to be $\le 4\%$. Its precise value depends 
on the mass of CDM particles (LSP) and the structure of the target. 
\end{abstract}

\vspace{1.7cm}


\vspace{3.0cm}

\end{titlepage}


\section{ Introduction }

There are many arguments supporting the fact that, the cold dark matter 
of the universe, i.e. its component which is composed of particles which 
were non-relativistic at the time of structure formation, is at least
$60\%$.~\cite{COBE} There are two interesting cold dark
matter candidates:
i) Massive Compact Halo Objects (MACHO's) and
ii) Weak Interacting Massive Particles (WIMP's).
The MACHO's cannot exceed $40\%$ of the CDM component.~\cite{Benne,Jungm}
In the present work we discuss a special WIMP candidate connected with
the supersymmetry, i.e. the lightest supersymmetric particle (LSP).

We examine the possibility to directly detect the LSP~\cite{JDV}-\cite{Haber}
via the recoiling
of a nucleus (A,Z) in the elastic scattering process:
\begin{equation}
\chi \, +\, (A,Z) \, \to \, \chi \,  + \, (A,Z)^* 
\end{equation}
($\chi$ denotes the LSP). In this investigation,
we proceed with the following steps:

1) We write down the effective Lagrangian at the elementary particle 
(quark) level obtained in the framework of supersymmetry as described 
in Refs.~\cite{Jungm}-\cite{Bottin}

2) We go from the quark to the nucleon level using an appropriate quark 
model for the nucleon. Special attention in this step is paid to the 
scalar couplings, which dominate the coherent part of the cross section
and the isoscalar axial current, which,
as we will see, strongly depend on the assumed quark model.~\cite{Dree}

3) We compute the relevant nuclear matrix elements~\cite{Ress}-\cite{Nikol}
using as reliable as possible many body nuclear wave functions hoping
that, by putting as accurate nuclear physics input as possible, 
one will be able to constrain the SUSY parameters as much as possible.

4) We calculate the modulation of the cross sections due to the earth's
revolution around the sun by a folding procedure
assuming a Maxwell Boltzmann distribution~\cite{Jungm} of velocities for LSP.

There are many popular targets~\cite{Smith}-\cite{Kane} 
for LSP detection as e.g. $^{19}F$,
$^{23}Na$, $^{27}Al$, $^{29}Si$, $^{40}Ca$, $^{73,74}Ge$, $^{127}I$, 
$^{207}Pb$, etc.
Among them $^{207}Pb$ has been recently proposed~\cite{KVprd} as 
a theoretical laboratory. Furthermore, it can be an important detector, 
since its spin matrix element, especially the isoscalar one, does 
not exhibit large quenching as that of the light and up to now much
studied $^{29}Si$ and $^{73}Ge$ nuclei.~\cite{Ress}

Our purpose is to calculate LSP-nucleus scattering cross section using 
some representative input in the restricted SUSY parameter 
space,~\cite{Bottin}-\cite{Roszko,Kane,Casta}
to compute the coherent LSP-nucleus scattering cross sections throughout 
the periodic table and study the spin matrix elements of $^{207}Pb$, 
since this target, in addition to its 
experimental qualifications, has the advantage of a rather simple nuclear
structure. We compare our results to those obtained~\cite{Ress} for other
proposed cold dark matter detection targets. We finally present results 
obtained by using new input SUSY parameters~\cite{Casta} obtained in a
phenomenologically allowed parameter space.

\section{Effective Lagrangian }

Before proceeding with the
construction of the effective Lagrangian we will briefly discuss 
the nature of the
lightest supersymmetric particle (LSP) focusing on those ingredients
which are of interest to dark matter.

\subsection{ The nature of LSP } 

In currently favorable supergravity models the LSP is a linear
combination~\cite{Jungm,JDV} of the neutral four fermions 
${\tilde B}, {\tilde W}_3, {\tilde H}_1$ and ${\tilde H}_2$ 
which are the supersymmetric partners of the gauge bosons $B_\mu$ and
$W^3_\mu$ and the Higgs scalars
$H_1$ and $H_2$. Admixtures of s-neutrinos are expected to be negligible.

In the above basis the mass-matrix takes the form~\cite{Jungm,Hab-Ka} 
\beq
  \left(\begin{array}{cccc}M_1 & 0 & -m_z c_\beta s_w & m_z s_\beta s_w \\
 0 & M_2 & m_z c_\beta c_w & -m_z s_\beta c_w \\
-m_z c_\beta s_w & m_z c_\beta c_w & 0 & -\mu \\
m_z s_\beta s_w & -m_z c_\beta c_w & -\mu & 0 \end{array}\right)
\label{eq:eg 4}
\eeq

In the above expressions $c_W = cos \theta_W$,  
$s_W = sin\theta_W$, $c_\beta = cos\beta$, $s_\beta = sin\beta$,
where $tan\beta =\langle \upsilon_2\rangle/\langle\upsilon_1\rangle$ is the
ratio of the vacuum expectation values of the Higgs scalars $H_2$ and
$H_1$. $\mu$ is a dimensionful coupling constant which is not specified by the
theory (not even its sign). The parameters $tan \beta , M_1, M_2, \mu$ are
determined by the procedure of Refs.~\cite{Kane,Casta} using universal
masses of the GUT scale.

By diagonalizing the above matrix we obtain a set of eigenvalues $m_j$ and
the diagonalizing matrix $C_{ij}$ as follows

\beq
\left(\begin{array}{c}{\tilde B}_R\\
{\tilde W}_{3R}\\
{\tilde H}_{1R} \\
 {\tilde H}_{2R}\end{array}\right) = (C_{ij})
\left(\begin{array}{c} \chi_{1R} \\
\chi_{2R} \\   \chi_{3R} \\
\chi_{4R} \end{array}\right)  \qquad
\left(\begin{array}{c} {\tilde B}_L \\
{\tilde W}_{2L} \\   {\tilde H}_{1L} \\
{\tilde H}_{2L} \end{array}\right) =  \left(C^*_{ij} \right)
\left(\begin{array}{c} \chi_{1L} \\
\chi_{2L} \\   \chi_{3L} \\
\chi_{4L} \end{array}\right) 
\label{eq:eg.8}
\eeq

\noindent
Another possibility to express the above results in photino-zino 
basis ${\tilde \gamma}, {\tilde Z}$ via
\barr
{\tilde W}_3 &=& sin \theta_W {\tilde \gamma}
 -cos \theta_W {\tilde Z} \nonumber \\
{\tilde B}_0 &=& cos \theta_W {\tilde \gamma}
 +sin \theta_W {\tilde Z} \label{eq:eg9}
\earr
In the absence of supersymmetry breaking $(M_1=M_2=M$ and $\mu=0)$ the
photino is one of the eigenstates with mass $M$. One of the remaining
eigenstates has a zero eigenvalue and is a linear combination of ${\tilde
H}_1$ and ${\tilde H}_2$ with mixing angle $sin \beta$. In the presence
of SUSY breaking terms the ${\tilde B}, {\tilde W}_3$ basis is superior
since the lowest eigenstate $\chi_1$ or LSP is primarily ${\tilde B}$. From
our point of view the most important parameters are the mass $m_x$ of LSP
and the mixings $C_{j1}, j=1,2,3,4$ which yield the $\chi_1$ content of the
initial basis states. These parameters which are relevant here
are shown in Table 1.

We are now in a position to find the interaction of $\chi_1$ with matter. 
We distinguish three possibilities involving Z-exchange, s-quark exchange and
Higgs exchange.

\subsection{ The relevant Feynman diagrams } 

\subsubsection{The Z-exchange contribution }

This can arise from the interaction of Higgsinos with $Z$ which can be read
from Eq. C86 of Ref.~\cite{Hab-Ka} 
\beq
{\it L} = \frac{g}{cos \theta_W} \frac{1}{4} [{\tilde H}_{1R}
\gamma_{\mu}{\tilde H}_{1R} -{\tilde H}_{1L}\gamma_{\mu} {\tilde H}_{1L} -
({\tilde H}_{2R}\gamma_{\mu}{\tilde H}_{2R}  -{\tilde H}_{2L}\gamma_{\mu}{\tilde
H}_{2L})]Z^{\mu}
 \label{eq:eg 10}
\eeq
Using Eq. (\ref{eq:eg.8}) and the fact that for Majorana particles 
${\bar \chi} \gamma_{\mu} \chi = 0$, we obtain 
\beq
{\it L} = \frac {g}{cos \theta_W} \frac {1}{4} (|C_{31}|^2
-|C_{41}|^2) {\bar \chi}_1\gamma_{\mu} \gamma_5 \chi_1 Z^{\mu}
 \label{eq:eg 11}
\eeq
which leads to the effective 4-fermion interaction (see Fig. 1)
\beq
{\it L}_{eff} = \frac {g}{cos \theta_W} \frac {1}{4} 2(|C_{31}|^2
-|C_{41}|^2)  (- \frac {g}{2cos \theta_W} \frac {1}{q^2 -m^2_Z}
 {\bar \chi}_1\gamma^{\mu} \gamma_5 \chi_1)J^Z_\mu
 \label{eq:eg 12}
\eeq
where the extra factor of 2 comes from the Majorana nature of
$\chi_1$. The neutral hadronic current $J^Z_\lambda$ is given by
\beq
J^Z_{\lambda} = - {\bar q} \gamma_{\lambda} \{ \frac {1}{3} sin^2 \theta_W -
\Big[ \,\frac {1}{2} (1-\gamma_5) - sin^2 \theta_W \Big]\tau_3 \} q  
  \label{eq:eg 13}
\eeq
at the nucleon level it can be written as
\beq
{\tilde J}_{\lambda}^Z = -{\bar N} \gamma_{\lambda} \{ \, sin^2 \theta_W 
-g_V (\frac{1}{2} - sin^2\theta_W)\tau_3 + \frac{1}{2} g_A \gamma_5 \tau_3
\} N
  \label{eq:eg 14}
\eeq
Thus we can write
\beq
{\it L}_{eff} = - \frac {G_F}{\sqrt 2} ({\bar \chi}_1 \gamma^{\lambda}
\gamma^5 \chi_1) J_{\lambda}(Z)
 \label{eq:eg 15}
\eeq 
where
\beq
J_{\lambda}(Z) = {\bar N} \gamma_{\lambda} [f^0_V(Z) + f^1_V(Z) \tau_3
+  f^0_A(Z) \gamma_5 + f^1_A(Z) \gamma_5  \tau_3] N
\label{eq:eg 16}
\eeq
and
\barr
f^0_V(Z) &=& 2(|C_{31}|^2 -|C_{41}|^2) \frac {m^2_Z}{m^2_Z - q^2} sin^2
\theta_W \\ 
f^1_V(Z) &=& - 2(|C_{31}|^2 -|C_{41}|^2) \frac
{m^2_Z}{m^2_Z - q^2}g_V (\frac {1}{2} - sin^2 \theta_W) \\
f^0_A (Z) &=& 0  \\
f^1_A (Z) &=&  2(|C_{31}|^2 -|C_{41}|^2) \frac {m^2_Z}{m^2_Z - q^2} \,\, \frac
{1}{2} g_A \label{eq:eg 13d}
\earr
with $g_V=1.0,  g_A = 1.24$. We can easily see that
\beq
f^1_{V}(Z)/ f^0_{V}(Z) = -g_V ( \frac {1}{2sin^2 \theta_W} - 1 ) \simeq
- 1.15  \nonumber
\eeq
Note that the suppression of this Z-exchange interaction compared to 
the ordinary
neutral current interactions arises from the smallness of the mixings
$C_{31}$ and $C_{41}$, a consequence of the fact that the Higgsinos are
normally quite a bit heavier than the gauginos.  Furthermore, the two
Higgsinos tend to cancel each other.
\vspace{0.2cm}

\subsubsection{ The $s$-quark mediated interaction }

The other interesting possibility arises from the other two components of
$\chi_1$, namely ${\tilde B}$ and ${\tilde W}_3$. Their corresponding
couplings to $s$-quarks can be read from the appendix C4 of Ref.~\cite{Hab-Ka}
They are
\barr
{\it L}_{eff} &=& -g \sqrt {2} \{{\bar q}_L [T_3 {\tilde W}_{3R} 
- tan \theta_W (T_3 -Q) {\tilde B}_R ] {\tilde q}_L \nonumber \\
&-& tan  \theta_W {\bar q}_R Q {\tilde B}_L {\tilde q}_R\} + HC
 \label{eq:eg 17}
\earr
where ${\tilde q}$ are the scalar quarks (SUSY partners of quarks). A
summation over all quark flavors is understood. Using Eq. (\ref{eq:eg.8}) we
can write the above equation in the $\chi_i$ basis. Of interest to us here
is the part
\barr 
{\it L}_{eff} &=& g \sqrt {2} \{(tan \theta_W (T_3 -Q) C_{11} -
T_3 C_{21}) {\tilde q}_L \chi_{1R} {\tilde q}_L \nonumber \\
&+& tan  \theta_W C_{11} Q {\bar q}_R \chi_{1L} {\tilde q}_R\} 
\label{eq:eg.18}
\earr
The above interaction is almost diagonal in the quark flavor. There exists,
however, mixing between the s-quarks ${\tilde q}_L$ and ${\tilde q}_R$
(of the same flavor) i.e.
\begin{equation}
{\tilde q}_L = cos \theta_{{\tilde q}}{\tilde q}_1 
+ sin \theta_{{\tilde q}}{\tilde q}_2   
\end{equation}
\begin{equation}
{\tilde q}_R = -sin \theta_{{\tilde q}}{\tilde q}_1 
+ cos \theta_{{\tilde q}}{\tilde q}_2   
\end{equation}
with
\begin{equation}
tan 2\theta_{{\tilde u}} =  \frac {m_u(A+\mu cot \beta)} 
{m^2_{u_L} -m^2_{{\tilde u}_R} + m^2_z cos2 \beta/2}
\end{equation}
\begin{equation}
tan 2\theta_{{\tilde d}} =  \frac {m_d(A+\mu tan \beta)} 
{m^2_{d_R} -m^2_{{\tilde d}_R} + m^2_Z cos2 \beta/2}
\end{equation}
Thus Eq. (\ref{eq:eg.18}) becomes
\barr
{\it L}_{eff} &=& g \sqrt{2} \left\{ [ B_L cos\theta_{{\tilde q}} \right.
{\bar q}_L \chi_{1R} -B_R sin\theta_{{\tilde q}}{\bar q}_R \chi_{1L}]
{\tilde q}_1 \nonumber \\
&+& [ B_L sin \theta_{{\tilde q}} {\bar q}_L \chi_{1R} + 
B_R cos\theta_{{\tilde q}} {\bar q}_R \chi_{1L}] \left.
{\tilde q}_2 \right \} \nonumber   
\earr
with
\begin{center}
$B_L(q) = -\frac{1}{6} C_{11}tan\theta_{\omega}-\frac{1}{2} C_{21},
\,\,\,  q=u \ \ (charge \ \ 2/3)$   
\end{center}
\begin{center}
$B_L(q) = -\frac{1}{6} C_{11} tan\theta_{\omega} + \frac{1}{2} C_{21},
\ \ \ q=d \ \ (charge \ \ -1/3)$   
\end{center}
\begin{center}
$B_R(q) = \frac {2}{3} tan \theta_{\omega}  C_{11}, \ \ \  q=u \ \ (charge \ \ 2/3)$   
\end{center}
\begin{center}
$B_R(q) = - \frac {1}{3} tan \theta_{\omega}  C_{11}, \ \ \  q=d\ \ (charge \ \ -1/3)$   
\end{center}
The effective four fermion interaction, Fig. 1, takes the form
\barr
\lefteqn{{\it L}_{eff} = (g \sqrt{2})^2  \{(B_L cos \theta_{\tilde q} {\bar q}_L
\chi_{1R} - B_R sin\theta_{\tilde q} {\bar q}_R \chi_{1L}) }      
 & & \nonumber \\
& & \frac{1}{q^2-m {\tilde q}_1^2} (B_L cos \theta_q {\bar \chi}_{1R} q_L
 - B_R sin\theta_{{\tilde q}} {\bar \chi}_{1L} q_R) \nonumber \\
 & & + (B_L sin \theta_q q_L
\chi_{1R} + cos\theta_{\tilde q} {\bar q}_R \chi_{1L}) \frac{1}{q^2-m 
{\tilde q}_2^2} \nonumber \\
 & &(B_L sin \theta_q {\bar \chi}_{1R} q_L
 + B_R cos\theta_{\tilde q} {\bar \chi}_{1L} q_R) \} \label{eq:eg 18}
\earr

The above effective interaction can be written as
\begin{equation}
{\it L}_{eff} = {\it L}_{eff}^{LL+RR} + {\it L}_{eff}^{LR}
\end{equation}
The first term involves quarks of the same chirality and is not much effected
by the mixing (provided that it is small). The second term involves quarks 
of opposite chirality and is proportional to the s-quark mixing.

\bigskip

i) The part ${\it L}_{eff}^{LL+RR}$

\bigskip

Employing a Fierz transformation ${\it L}_{eff}^{LL+RR}$ can be cast in the more
convenient form
 \barr 
{\it L}_{eff}^{LL+RR} =&& (g \sqrt{2})^2  2(-\frac{1}{2})\{ |B_L|^2 
\nonumber  \\
&& (\frac{cos^2 \theta_{\tilde q}}{q^2-m {\tilde q}_1^2} + 
\frac{sin^2\theta_{\tilde q}}{q^2-m {\tilde q}_2^2} )
 {\bar q}_L \gamma_\lambda q_L
\chi_{1R} \gamma^\lambda \chi_{1R} 
\nonumber  \\
&& + |B_R|^2 (\frac {sin^2 \theta_{\tilde q}}{q^2-m {\tilde q}_1^2} + 
\frac {cos^2 {\theta_{\tilde q}}} {q^2-m {\tilde q}_2^2} )
 {\bar q}_R \gamma_\lambda q_R
\chi_{1L} \gamma^\lambda \chi_{1L} \} 
 \label{eq:eg.19}
\earr
The factor of 2 comes from the majorana nature of LSP and the (-1/2) comes
from the Fierz transformation. Equation (\ref{eq:eg.19}) can be
written more compactly as
\barr
{\it L}_{eff} & = &  - \frac {G_F} {\sqrt {2}} 2\{ {\bar q} \gamma_\lambda
(\beta_{0R} +\beta_{3R} \tau_3) (1+ \gamma_5) q
\nonumber \\
 & - & {\bar q} \gamma_\lambda (\beta_{0L} +\beta_{3L} \tau_3)(1-\gamma_5)
 q \}({\bar \chi}_1 \gamma^\lambda \gamma^5 \chi_1 \}
 \label{eq:eg 21}
\earr
with
\barr
\beta_{0R} &=& \Big( \frac {4} {9} \chi^2_{{\tilde u}_R} 
+\frac {1} {9} \chi^2_{{\tilde d}_R}\Big) |C_{11} tan \theta_W|^2\nonumber \\
\beta_{3R} &=& \Big( \frac {4} {9} \chi^2_{{\tilde u}_R} 
-\frac {1} {9} \chi^2_{{\tilde d}_R} \Big) |C_{11} tan \theta_W|^2
\label{eq:eg 22}\\ 
\beta_{0L} &=& | \frac {1} {6} C_{11} tan\theta_W
+\frac{1}{2} C_{21}|^2 \chi^2_{{\tilde u}_L} + | \frac {1} {6} C_{11}
tan\theta_W  - \frac{1}{2} C_{21}|^2 \chi^2_{{\tilde d}_L} \nonumber \\
\beta_{3L} &=& | \frac {1} {6} C_{11} tan\theta_W +\frac{1}{2} C_{21}|^2
\chi^2_{{\tilde u}_L} - | \frac {1} {6} C_{11} tan\theta_W 
-\frac{1}{2} C_{21}|^2 \chi^2_{{\tilde d}_L} \nonumber 
\earr
with
\barr
\chi^2_{qL} &=& c^2_{\tilde q} \frac {m_W^2}{m_{{\tilde q}^2_1}-q^2} +
 s^2_{{\tilde q}} \frac {m_W^2}{m_{{\tilde q}^2_2}-q^2} \nonumber \\
\chi^2_{qR} &=& s^2_{\tilde q} \frac{m_W^2}{m_{{\tilde q}^2_1}-q^2} +
 c^2_{\tilde q} \frac{m_W^2}{m_{{\tilde q}^2_2}-q^2} \nonumber \\
c_{\tilde q} &=& cos\theta_{\tilde q}, \,\ s_{\tilde q}=sin\theta_{\tilde q} 
 \label{eq:eg 23}
\earr
The above parameters are functions of the four-momentum transfer which
in our case is negligible. Proceeding as in Sec. 2.2.1 we can obtain the
effective Lagrangian at the nucleon level as
\beq
{\it L}_{eff}^{LL+RR} = - \frac {G_F}{\sqrt 2} ({\bar \chi}_1 \gamma^{\lambda}
\gamma^5 \chi_1) J_{\lambda} ({\tilde q})
 \label{eq:eg.24}
\eeq
\beq
J_{\lambda}({\tilde q}) = {\bar N} \gamma_{\lambda} \{f^0_V({\tilde q}) + f^1_V
({\tilde q}) \tau_3 + f^0_A({\tilde q}) \gamma_5 + f^1_A({\tilde q}) \gamma_5
\tau_3) N
  \label{eq:eg.25}
\eeq
with
\barr
 f^0_V = 6(\beta_{0R}-\beta_{0L}) , \qquad f^1_V = 2
(\beta_{3R}-\beta_{3L})
\nonumber \\
f^0_A = 2g_V (\beta_{0R}+\beta_{0L}), \qquad
f^1_A = 2g_A(\beta_{3R}+\beta_{3L})
\label{eq:eg 25a}
\earr

We should note that this interaction is more suppressed than the ordinary
weak interaction by the fact that the masses of the s-quarks are usually
larger than that of the gauge boson $Z^0$. In the limit in which the LSP 
is a pure bino ($C_{11} = 1,  C_{21} = 0$) we obtain
\barr
\beta_{0R} &=& tan^2 \theta_W \Big( \frac{4}{9} \chi^2_{u_R} 
+\frac{1} {9} \chi^2_{{\tilde d}_R}\Big) \nonumber \\
\beta_{3R} &=& tan^2 \theta_W \Big( \frac{4} {9} \chi^2_{u_R} 
-\frac {1} {9} \chi^2_{{\tilde d}_R} \Big) 
\nonumber  \\
\beta_{0L} &=&  \frac{tan^2 \theta_W} {36} (\chi^2_{{\tilde u}_L} + 
 \chi^2_{{\tilde d}_L}) \nonumber \\
\beta_{3L} &=&  \frac{tan^2 \theta_W} {36}
(\chi^2_{{\tilde u}_L} -  \chi^2_{{\tilde d}_L})
\label{eq:eg 24}
\earr
Assuming further that $\chi_{{\tilde u}_R} = \chi_{{\tilde d}_R} 
= \chi_{{\tilde u}_L} = \chi_{{\tilde d}_L}$ we obtain
\barr
 f^1_V ({\tilde q}) / f^0_V({\tilde q}) &\simeq& + \frac{2}{9}
\nonumber \\
  f^1_A ({\tilde q})/ f^0_A({\tilde q}) &\simeq& + \frac{6}{11}
 \label{eq:eg 25}
\earr

If, on the other hand, the LSP were the photino ($C_{11} = cos\theta_W,
C_{21} = sin \theta_W, C_{31} = C_{41} = 0$) and the s-quarks were
degenerate there would be no coherent contribution ($f^0_V = 0$ if
$\beta_{0L} =\beta_{0R}$).

\bigskip

ii) ${\it L}_{eff}^{LR}$

\bigskip

From Eq. (\ref{eq:eg 18}) we obtain

$$
{\it L}_{eff}^{LR} = - (g \sqrt{2})^2 sin2\theta_{\tilde q} B_L(q) B_R(q)
\frac{1}{2} [ \frac {1} {q^2-m {\tilde q}_1^2} - 
\frac{1}{q^2-m {\tilde q}_2^2} ]
$$
$$
({\bar q}_L \chi_{1R} {\bar \chi}_{1L} q_R + 
{\bar q}_R \chi_{1L} {\bar \chi}_{1R} q_L)
$$
Employing a Fierz transformation we can cast it in the form
\barr
\lefteqn{{\it L}_{eff} = - \frac{G_F} {\sqrt{2}} [\beta_+ ({\bar q} q 
{\bar \chi}_1 \chi_1
+ {\bar q}\gamma_5 q {\bar \chi}_1 \gamma_5\chi_1 -({\bar q}\sigma_{\mu\nu} q)
({\bar \chi}_1 \sigma^{\mu\nu}\chi_1))} & & \nonumber \\
 & &+ \beta_- ({\bar q} \tau_3 q {\bar \chi}_1 \chi_1
+ {\bar q} \tau_3 \gamma_5 q {\bar \chi}_1 \gamma_5 \chi_1 - {\bar q} 
\sigma_{\mu\nu} \tau_3 q {\bar \chi}_1 \sigma^{\mu\nu} \chi_1)] \nonumber   
\earr
where
\barr
\beta_{\pm} &=& \frac{1}{3} tan\theta_W C_{11} \{ 2sin 2\theta_{\tilde u}
[\frac{1}{6} C_{11} tan\theta_W + \frac{1}{2} C_{21}] \Delta_{\tilde u}
 \nonumber \\
 &\mp& sin 2 \theta_{\tilde d}
[\frac{1}{6} C_{11} tan\theta_W -  \frac{1}{2} C_{21}] \Delta_{\tilde d} \}  
\nonumber
\earr
with
\begin{center}
$\Delta_{\tilde u} = \frac{(m^2_{{\tilde u}_1}-m^2_{{\tilde u}_2}) M^2_W}
{(m^2_{{\tilde u}_1}-q^2)(m^2_{{\tilde u}_2}-q^2)}$
\end{center}
and an analogous equation for $\Delta_{\tilde d}$. Here $u$ indicate quarks with
charge 2/3 and d quarks with charge -1/3.

In going to the nucleon level and ignoring the negligible pseudoscalar and
tensor components in the spirit of Ref.~\cite{Adler} we obtain
\beq
{\it L}_{eff} = \frac{G_F} {\sqrt {2}} [f^0_s ({\tilde q}) {\bar N} N +
 f^1_s ({\tilde q}) {\bar N} \tau_3 N] {\bar \chi}_1 \chi_1
\eeq
with
\beq
f^0_s ({\tilde q}) = 1.86 \beta_+  \ \ \  and \ \ \
 f^1_s ({\tilde q}) = 0.48 \beta_- \ \ (Model A)
\eeq
(see sect. 2.2.3)
The appearance of scalar terms in s-quark exchange has been first noticed
in Ref.~\cite{Gries} It has also been noticed there that one should consider
explicitly the effects of quarks other than u and d~\cite{Dree} in going 
from the quark to the nucleon level. We first notice that with the exception 
of $t$ s-quark the ${\tilde q}_L - {\tilde q}_R$ mixing small. Thus

\barr
sin 2\theta_{\tilde u} \Delta {\tilde u} &\simeq& \frac{2m_u(A + \mu cot\beta)
m^2_W}  {(m^2_{{\tilde u}_L}-q^2) (m^2_{{\tilde u}_R}-q^2)}\nonumber \\
sin 2\theta_{\tilde d} \Delta {\tilde d} &\simeq& \frac{2m_d(A + \mu tan\beta)
m^2_W}  {(m^2_{{\tilde d}_L}-q^2) (m^2_{{\tilde d}_R}-q^2)}\nonumber
\earr
Then the amplitude for this s-quark contribution is proportional to the 
quark mass (${\grave a}$ la Higgs). Thus the amplitude for finding such 
quarks in the nucleon can be computed in a way which is similar to that of
the Higgs coupling (see Sec. 2.2.3). For the t s-quark the mixing is complete,
which implies that the amplitude is independent of the top quark mass.
Hence in the case of the top quark we do not get an extra enhancement in
going from the quark to the nucleon level. As we will see in the next
section we get an enhancement due to  quarks other than $u$ and $d$ (see
model B in the next section). This is not enough, however, to dominate even 
over $\beta f^0_V$ in the SUSY parameter space considered here.
Thus, $f^0_S({\tilde q})$ can be neglected in front of the isoscalar
scalar coupling coming from Higgs exchange (see sect. 2.2.3).
 
\subsubsection{ The intermediate Higgs contribution}

The coherent scattering can be mediated via the intermediate Higgs
particles which survive as physical particles (see Fig. 2).
The relevant interaction can arise out of the
Higgs-Higgsino-gaugino interaction which takes the form
\barr
 {\it L}_{H \chi \chi} &=& \frac {g}{\sqrt 2} \Big({\bar{\tilde W}}^3_R
 {\tilde H}_{2L} H^{0*}_2 - {\bar{\tilde W}}^3_R {\tilde H}_{1L} H^{0*}_1 
\nonumber \\
   &-&tan \theta_w ({\bar{\tilde B}}_R
  {\tilde H}_{2L} H^{0*}_2 - {\bar{\tilde B}}_R {\tilde H}_{1L} H^{0*}_1)
 \Big) + H.C.
 \label{eq:eg 26}
\earr
Proceeding as above we can express ${\tilde W}$ an ${\tilde B}$ in terms
of the appropriate eigenstates and retain the LSP to obtain
\barr
{\it L} &=& \frac {g}{\sqrt 2} \Big((C_{21} -tan\theta_w C_{11})
C_{41}{\bar \chi}_{1R} \chi_{1L} H^{o*}_2 \nonumber \\
  &-&(C_{21} -tan \theta_w C_{11}) 
C_{31}{\bar \chi}_{1R} \chi_{1L} H^{o*}_1 \Big) + H.C.
\label{eq:eg 27}
\earr

We can now proceed further and express the fields 
${H^0_1}^*$, ${H^0_2}^*$ in terms of the physical fields $h$, $H$ and
$A.$ The term which contains $A$ will be neglected, since it yields only
a pseudoscalar coupling which does not lead to coherence.

Thus we can write
\beq
{\cal L}_{eff} = - \frac{G_F}{\sqrt{2}}{\bar \chi} \chi \,
{\bar N} [ f^0_s (H) + f^1_s (H) \tau_3 ] N
\label{2.1.1}
\eeq
where
\beq
f^0_s (H)  = \frac{1}{2} (g_u + g_d) + g_s + g_c + g_b + g_t
\label{2.1.2}
\eeq
\beq
f^1_s (H)  = \frac{1}{2} (g_u - g_d)
\label{2.1.3}
\eeq
with
\beq
g_{a_i}  = \big[ g_1(h) \frac{cos \alpha}{sin \beta}
+ g_2(H) \frac{sin \alpha}{sin \beta} \big] \frac{m_{a_i}}{m_N},
\quad a_i = u,c,t
\label{2.1.4}
\eeq
\beq
g_{\kappa_i}  = \big[- g_1(h) \frac{sin \alpha}{cos \beta}
+ g_2(H) \frac{cos \alpha}{cos \beta} \big] 
\frac{m_{\kappa_i}}{m_N},
\quad \kappa_i = d,s,b
\label{2.1.5}
\eeq
\beq
g_{1}(h)  = 4 (C_{11 } tan \theta_W - C_{21}) (C_{41} cos \alpha - C_{31}
sin \alpha) \frac{m_N m_W}{m^2_h -q^2}
\label{2.1.6}
\eeq
\beq
g_{2}(H)  = 4 (C_{11 } tan \theta_W - C_{21}) (C_{41} sin \alpha - C_{31}
cos \alpha) \frac{m_N m_W}{m^2_H -q^2}
\label{2.1.7}
\eeq
where $m_N$ is the nucleon mass, and
the parameters $m_h$, $m_H$ and $\alpha$ depend on the SUSY parameter
space (see Table 1). If one ignores quarks other than
$u$ and $d$ (model A) and uses $m_u =5 MeV=m_d/2$ finds~\cite{Adler}
\beq
f^0_s = 1.86 (g_u + g_d)/2, \quad
f^1_s = 0.49 (g_u - g_d)/2, \quad
\label{2.1.8}
\eeq

As we have already mentioned, one has to be a bit more 
careful in handling quarks other than $u$ and $d$ since their couplings
are proportional to their mass.~\cite{Dree,Bottin}
One encounters in the nucleon not only
sea quarks ($u {\bar u}, d {\bar d}$ and $s {\bar s}$) but the heavier
quarks also due to QCD effects, which were estimated at
the one loop level in Ref.~\cite{Chen88,Chen89} This way one obtains
the pseudoscalar Higgs-nucleon
coupling $f^0_s(H)$ by using  effective quark masses as follows
\begin{center}
$m_u \ra f_u m_N, \ \ m_d \ra f_d m_N. \ \ \  m_s \ra f_s m_N$   
\end{center}
\begin{center}
$m_Q \ra f_Q m_N, \ \ (heavy\ \  quarks \ \ c,b,t)$   
\end{center}
where $m_N$ is the nucleon mass. The isovector contribution is now
negligible. The parameters $f$ can be obtained by chiral symmetry breaking 
terms in relation to phase shift and dispersion analysis.~\cite{Dree,Bottin}
Following Cheng~\cite{Chen88,Chen89} we obtain
\begin{center}
$ f_u = 0.021, \quad f_d = 0.037, \quad  f_s = 0.140, 
\quad \sum_Q f_Q = 0.240$ \quad  (model B)   
\end{center}
Another possible solution is
\begin{center}
$ f_u = 0.023, \quad f_d = 0.034, \quad  f_s = 0.400, \quad
\sum_Q f_Q = 0.120$  \quad (model C)   
\end{center}
In the present work we will consider these solutions (models B and C) and
compare them with the solution obtained in the spirit of Addler 
{\it et al.}~\cite{Adler} (Model A above).
For a more detailed discussion we refer the reader to 
Refs.~\cite{Dree,Bottin}

\subsection{ Expressions for the nuclear matrix elements } 

Combining for results of the previous section we can write
 \beq
{\it L}_{eff} = - \frac {G_F}{\sqrt 2} \{({\bar \chi}_1 \gamma^{\lambda}
\gamma_5 \chi_1) J_{\lambda} + ({\bar \chi}_1 
 \chi_1) J\}
 \label{eq:eg 41}
\eeq
where
\beq
  J_{\lambda} =  {\bar N} \gamma_{\lambda} (f^0_V +f^1_V \tau_3
+ f^0_A\gamma_5 + f^1_A\gamma_5 \tau_3)N
 \label{eq:eg.42}
\eeq
with
\barr
 f^0_V  &=&  f^0_V(Z) + f^0_V({\tilde q}),  \qquad
f^1_V  =  f^1_V(Z) + f^1_V({\tilde q})
\nonumber \\
  f^0_A  &=&  f^0_A(Z) + f^0_A({\tilde q}) , \qquad
f^1_A  =  f^1_A(Z) + f^1_A({\tilde q})
 \label{eq:eg 43}
\earr
and
\beq
J = {\bar N} (f^0_s +f^1_s \tau_3) N
 \label{eq:eg.45}
\eeq

We have neglected the uninteresting pseudoscalar and tensor
currents. Note that, due to the Majorana nature of the LSP, 
${\bar \chi_1} \gamma^{\lambda} \chi_1 =0$ (identically).
We have seen that, the vector and axial vector form factors can
arise out of Z-exchange and s-quark exchange.~\cite{Goodm}-\cite{JDV}
They have uncertainties in them. Here we consider the three choices in the
allowed parameter space of Ref.~\cite{Kane} 
and the eight parameter choices of Ref.~\cite{Casta}
These involve universal soft breaking masses at the scale. Non-universal
masses have also recently been employed.~\cite{Dree}-\cite{Arnow}
In our choice of the parameters the LSP is mostly a gaugino. Thus, the Z-
contribution is small. It may become dominant in models in which the LSP
happens to be primarily a Higgsino.~\cite{KA-JA} 
The transition from the quark to the nucleon level is pretty straightforward 
in the case of vector current contribution. 
We will see later that, due to the Majorana nature of the LSP, the
contribution of the vector current, which can lead to a coherent effect of all
nucleons, is suppressed.~\cite{JDV} 
The vector current is effectively multiplied by a factor of $\beta=v/c$, 
$v$ is the velocity of LSP (see Tables 2(a),(b)).
Thus, the axial current, especially in the
case of light and medium mass nuclei, cannot be ignored.

For the isovector axial current one is pretty confident about how to go 
from the quark to the nucleon level. We know from ordinary weak decays
that the coupling merely gets renormalized from $g_A=1$ to $g_A=1.24$.
For the isoscalar axial current the situation is not completely clear.
The naive quark model (NQM) would give a renormalization parameter of unity
(the same as the isovector vector current). This point of view has, 
however, changed in recent years due to the so-called spin
crisis,~\cite{Ashm}-\cite{Gensin} i.e.
the fact that in the EMC data~\cite{Ashm} it appears that only a small fraction
of the proton spin arises from the quarks. Thus, one may have to
renormalize $f^0_A$ by $g^0_A=0.28$, for u and d quarks, and $g^0_A=-0.16$
for the strange quarks,~\cite{Jaffe,Gensin} i.e. a total factor of 0.12. 
These two possibilities, labeled as NQM and EMC, are listed in 
Tables 2(a),(b). One cannot completely rule out the possibility that the 
actual value maybe anywhere in the above mentioned region.~\cite{Gensin}

The scalar form factors arise out of the Higgs exchange or via s-quark 
exchange when there is mixing~\cite{Dree} between s-quarks ${\tilde q}_L$
and ${\tilde q}_R$
(the partners of the left-handed and right-handed quarks).
We have seen in Ref.~\cite{JDV} that they have two types 
of uncertainties in them. One, which is the most important,
at the quark level due to the uncertainties in the Higgs sector. 
The actual values of the parameters $f^0_S$ and $f^1_S$ used here, arising 
mainly from Higgs exchange, were obtained by considering 1-loop corrections
in the Higgs sector. As a result, the lightest Higgs mass is now a bit higher,
i.e. more massive than the value of the Z-boson.~\cite{El-Ri,Haber}

The other type of uncertainty is related to the step going from  
the quark to the nucleon level~\cite{Dree} (see sect. 2.2.3).
Such couplings are proportional to the
quark masses, and hence sensitive to the small admixtures of $q {\bar q }$
(q other than u and d) present in the nucleon. Again values of $f^0_S$ and
$f^1_S$ in the allowed SUSY parameter space are considered (see Tables
 2(a),(b)).

The invariant amplitude in the case of non-relativistic LSP 
can now be cast in the form~\cite{JDV}
\barr
|{\cal M}|^2 &=& \frac{E_f E_i -m^2_x +{\bf p}_i\cdot {\bf p}_f } {m^2_x} \,
|J_0|^2 +  |{\bf J}|^2 +  |J|^2 
 \nonumber \\ & \simeq & \beta ^2 |J_0|^2 + |{\bf J}|^2 + |J|^2 
\label{2.1}
 \earr
where $m_x$ is the LSP mass, $|J_0|$ and $|{\bf J }|$ indicate the matrix 
elements of the time and space components of the current $J_\lambda$ 
of Eq. (\ref{eq:eg.42}), respectively, and $J$ represents the matrix 
element of the 
scalar current J of Eq. (\ref{eq:eg.45}). Notice that $|J_0|^2$ is multiplied
by $\beta^2$ (the suppression due to the Majorana nature of LSP mentioned
above). It is straightforward to show that 
\beq
 |J_0|^2 = A^2 |F({\bf q}^2)|^2 \,\left(f^0_V -f^1_V \frac{A-2 Z}{A}
 \right)^2
\label{2.2}
\eeq
\beq
 J^2 = A^2 |F({\bf q}^2)|^2 \,\left(f^0_S -f^1_S \frac{A-2 Z}{A}
 \right)^2
\label{2.3}
\eeq
\beq
|{\bf J}|^2 = \frac{1}{2J_i+1} |\langle J_i ||\, [ f^0_A {\bf
\Omega}_0({\bf q})\,+\,f^1_A {\bf \Omega}_1({\bf q}) ] \,||J_i\rangle |^2 
\label{2.4}
\eeq
with $F({\bf q}^2)$ the nuclear form factor and
\beq 
{\bf \Omega}_0({\bf q})  = \sum_{j=1}^A {\bf \sigma}(j) e^{-i{\bf q} \cdot
{\bf x}_j }, \qquad
{\bf \Omega}_1({\bf q})  = \sum_{j=1}^A {\bf \sigma} (j) {\bf \tau}_3 (j)
 e^{-i{\bf q} \cdot {\bf x}_j }
\label{2.5}  
\eeq
where ${\bf \sigma} (j)$, ${\bf \tau}_3 (j)$, ${\bf x}_j$ are the spin, third
component of isospin ($\tau_3 |p\rangle = |p\rangle$) and coordinate of
the j-th nucleon and $\bf q$ is the momentum transferred to the nucleus.

The differential cross section in the laboratory frame takes the form~\cite{JDV}
\barr
\frac{d\sigma}{d \Omega} &=& \frac{\sigma_0}{\pi} (\frac{m_x}{m_N})^2
\frac{1}{(1+\eta)^2} \xi  \{\beta^2 |J_0|^2  [1 - \frac{2\eta+1}{(1+\eta)^2}
\xi^2 ] + |{\bf J}|^2 + |J|^2 \} 
\label{2.6}
 \earr
where $m_N$ is the proton mass, $\eta = m_x/m_N A$, $ $
$\xi = {\bf {\hat p}}_i \cdot {\bf {\hat q}} \ge 0$ (forward scattering) and  
\beq
\sigma_0 = \frac{1}{2\pi} (G_F m_N)^2 \simeq 0.77 \times 10^{-38}cm^2 
\label{2.7} 
\eeq
The momentum transfer  $\bf q$ is given by
\beq
|{\bf q}| = q_0 \xi, \qquad q_0 = \beta \frac{2  m_x c }{1 +\eta}
\label{2.8} 
\eeq
Some values of $q_0$ (forward momentum transfer) for some characteristic values
of $m_x$ and representative nuclear systems (light, medium and heavy)
are given in Table 3. It is clear from Eq. (\ref{2.8}) that the momentum
transfer can be sizable for large $m_x$ and heavy nuclei ($\eta$ small).

The total cross section can be cast in the form
\barr
\sigma &=& \sigma_0 (\frac{m_x}{m_N})^2 \frac{1}{(1+\eta)^2} \,
 \{ A^2 \, [[\beta^2 (f^0_V - f^1_V \frac{A-2 Z}{A})^2 
\nonumber \\ & + & 
(f^0_S - f^1_S \frac{A-2 Z}{A})^2 \, ]I_0(u_0) -
\frac{\beta^2}{2} \frac{2\eta +1}{(1+\eta)^2}
(f^0_V - f^1_V \frac{A-2 Z}{A})^2 I_1 (u_0) ]
\nonumber \\ & + & 
(f^0_A \Omega_0(0))^2 I_{00}(u_0) + 2f^0_A f^1_A \Omega_0(0) \Omega_1(0)
I_{01}(u_0)  \nonumber \\ 
&+& (f^1_A \Omega_1(0))^2 I_{11}(u_0) \, \} 
\label{2.9}
 \earr
The quantities $I_{\rho}$ entering Eq. (\ref{2.9}) are defined as 
(see Fig. 3)
\beq
I_\rho(u_0)  =   (1+\rho)u_0^{-(1+\rho)}
  \int_0^{u_0} x^{1+ \rho} \, |F( x)|^2 \,dx,
\qquad \rho = 0,1
\label{2.16} 
\eeq
where $F(q)$ the nuclear form factor and
\beq
u_0 = q_0^2b^2/2
\label{2.17} 
\eeq
Using appropriate expressions for the form factors (in a harmonic oscillator
basis with size parameter $b$) we obtain~\cite{KV92}-\cite{KV90}
\beq
I_\rho(u_0)  = \frac{1}{A^2} \{ \, Z^2 I^{(\rho)}_{ZZ}(u_0) +
2NZI^{(\rho)}_{NZ}(u_0) +N^2I^{(\rho)}_{NN}(u_0) \}
\label{2.18} 
\eeq
where
\barr
I^{(\rho)}_{\alpha \beta}(u_0)  =  (1 + \rho) 
&&\sum_{\lambda =0}^{N_{max}(\alpha)}  \, \sum_{\nu =0}^{N_{max}(\beta)} 
\frac{\theta_\lambda^{(\alpha)}}{\alpha} \, 
\frac{\theta_\nu^{(\beta)}}{\beta} \,
\frac{2^{\lambda +\nu+\rho} \,(\lambda +\nu+\rho)!}{u_0^{1+\rho}} \\
\nonumber 
&&\times \Big[ 1 - e^{-u_0} \sum_{\kappa =0}^{\lambda +\nu+\rho} \,\, 
\frac{u_0^\kappa}{\kappa!} \Big]
 \label{2.19} 
\earr
($\alpha,\beta = N, Z$).
The coefficients $\theta_\lambda^{(\alpha)}$ are given in Ref.,~\cite{KV92}
for light and medium nuclei, and in Ref.~\cite{KVcdm} for heavy nuclei.

The integrals $I_{\rho\rho^{\prime}}$, with $\rho,\rho^{\prime} =0,1$,
(see Fig. 4) result by following the standard procedure of the multipole
expansion of the $e^{-i {\bf q} \cdot {\bf r}}$ in Eq. (\ref{2.5}). One finds
\beq
I_{\rho \rho^{\prime}}(u_0) = 2 \int_0^1 \xi \, d\xi \sum_{\lambda,\kappa}
\frac{\Omega^{(\lambda,\kappa)}_\rho( u_0\xi^2)}{\Omega_\rho (0)} \,
\frac{\Omega^{(\lambda,\kappa)}_{\rho^{\prime}}( u_0\xi^2)}
{\Omega_{\rho^{\prime}}(0)} 
, \qquad \rho, \rho^{\prime} = 0,1
\label{2.11} 
\eeq
where, in the special case in which the ground state of $^{207}Pb$ is
approximated by a $2p_{1/2}$ neutron-hole, one finds
\beq
I_{00} = I_{01} = I_{11} = 2 \int_0^1 \xi \, [ F_{2p} (u_0 \xi^2) ]^2 \,d\xi
\label{2.21} 
\eeq
Even though the probability of finding a pure $2p_{1/2}$ neutron hole
in the $\frac{1}{2}^-$ ground state of $^{207}Pb$ is greater than 95\%,
the ground state magnetic moment is quenched due to the $1^+$ p-h excitation
involving the spin orbit partners. Hence, we expect a similar suppression
of the isovector spin matrix elements~\cite{Ver71}-\cite{Herli} (see
Table 4(a),(b)). For comparison, we present our results for $A=207$ together
with those of $A=29$ and $A=73$ (Ref.~\cite{Ress}) in Table 4(a).

\section{ Convolution of the cross section with the velocity distribution} 

The  cross sections which would be given from an LSP-detector participating 
in the revolution of the earth around the sun would appear retarded. In this
section we are going to study this effect by using the method of folding. To 
this aim let us assume that the LSP is moving with velocity $v_z$ with 
respect to the detecting apparatus. Then the detection rate 
for a target with mass $m$ is given by

\beq
\frac{dN}{dt} =\frac{\rho (0)}{m} \frac{m}{A m_N} | v_z | \sigma (v)
\label{4.1}  
\eeq
where $\rho (0) = 0.3 GeV/cm^3$ is the LSP density in our vicinity. 
This density has to be consistent with the LSP velocity distribution.
Such a consistent choice can be a Maxwell distribution~\cite{Jungm}
\beq
f(v^{\prime}) = (\sqrt{\pi}v_0)^{-3} e^{-(v^{\prime}/v_0)^2 }
\label{4.2}  
\eeq
provided that
\beq
v_0 = \sqrt{(2/3) \langle v^2 \rangle } =220 Km /s
\label{4.3}  
\eeq
For our purposes it is convenient to express the above distribution in the
laboratory frame, i.e.
\beq
f({\bf v}, {\bf v}_E) = (\sqrt{\pi}v_0)^{-3} 
e^{- ({\bf v}+{\bf v}_E)^2/v_0^2}
\label{4.4}  
\eeq
where ${\bf v}_E$ is the velocity of the earth with respect to the center
of the distribution. Choosing a coordinate system in which ${\bf \hat  x}_2$ 
is the axis of the galaxy, ${\bf \hat  x}_3$ is
along the sun's direction of motion (${\bf v}_0$)
and ${\bf \hat  x}_1 = {\bf \hat  x}_2 \times {\bf \hat  x}_3$, 
we find that the position of the axis of the ecliptic is determined
by the angle $\gamma \approx 29.80$ (galactic latitude) and the
azimuthal angle $\omega = 186.3^0$ measured on the galactic plane
from the ${\bf \hat  x}_3$ axis.~\cite{Alissan}

Thus, the axis of the ecliptic lies very close to the $x_2x_3$ plane
and the velocity of the earth is
\beq
{\bf v}_E \, = \, {\bf v}_0 \, + \, {\bf v}_1 \, 
= \, {\bf v}_0 + v_1(\, sin{\alpha} \, {\bf \hat x}_1
-cos {\alpha} \, cos{\gamma} \, {\bf \hat x}_2
+cos {\alpha} \, sin{\gamma} \, {\bf \hat x}_3\,)
\label{4.5}  
\eeq
and
\beq
{\bf v}_0 \cdot {\bf v}_1 = v_0 v_1 
\frac{cos\, \alpha}{\sqrt{1 + cot^2 \gamma \, cos^2\omega}}
\approx v_0 v_1 \, sin\, \gamma \, cos\,\alpha
\label{4.6}  
\eeq
where 
${ v}_0$ is the velocity of the sun around the center of the galaxy,
${ v}_1$ is the speed of the earth's revolution around the sun,
$\alpha$ is the phase of the earth orbital motion, $\alpha =2\pi 
(t-t_1)/T_E$, where $t_1$ is around second of June and
$T_E =1 year$.

The mean value of the event rate of Eq. (\ref{4.1}), is defined by
\beq
\Big<\frac{dN}{dt}\Big> =\frac{\rho (0)}{m} 
\frac{m}{A m_N} 
\int f({\bf v}, {\bf v}_E) \mid v_z \mid \sigma (|{\bf v}|)
d^3 {\bf v} 
\label{4.7}  
\eeq
Then we can write the counting rate as
\beq
\Big<\frac{dN}{dt}\Big> =\frac{\rho (0)}{m} \frac{m}{Am_N} \sqrt{\langle
v^2\rangle } \langle \Sigma\rangle 
\label{4.8}  
\eeq
where
\beq
\langle \Sigma\rangle =\int \frac{ | v_z | } {\sqrt{ \langle v^2 \rangle}} 
f({\bf v}, {\bf v}_E) \sigma (|{\bf v}|) d^3 {\bf v}
\label{4.9}  
\eeq
Thus, taking the polar axis in the direction ${\bf v}_E$, we get
\beq
\langle \Sigma \rangle = \frac{4}{\sqrt{6\pi} v_0^{4}}
\int_0^{\infty} v^3 d v \int_{-1}^{1} |\xi| d \xi 
e^{-(v^2+v_E^2+2v v_E \xi)/v_0^2 } \sigma (v) 
\label{4.10}  
\eeq
or
\beq
\langle \Sigma \rangle = \frac{2}{\sqrt{6\pi} v_E^{2}}
\int_0^{\infty} v d v \, F_0(\frac{2vv_E}{v^2_0}) \,
e^{-(v^2+v_E^2)/v_0^2 } \sigma (v) 
\label{4.11}  
\eeq
with 
\beq
F_0(\chi) =\chi sinh \chi - cosh \chi + 1
\label{4.12}  
\eeq

One can also write Eq. (\ref{4.11}) as follows
\beq
\langle \Sigma\rangle = \Big(\frac{2}{3} \Big)^{\frac{1}{2}}
\int_0^{\infty} \frac{v}{v_0} \, f_1(v) \sigma (v) dv
\label{4.13}  
\eeq
with
\beq
f_1(v) \, = \, \frac{1}{\sqrt{\pi}} \, \frac{v_0}{v^2_E}\, 
F_0(\frac{2vv_E}{v^2_0}) \, e^{-(v^2+v_E^2)/v_0^2 }
\label{4.14}  
\eeq
In the case in which the first term in Eq. (\ref{4.12}) becomes dominant,
we get 
\beq
f_1(v) \, = \, \frac{1}{\sqrt{\pi}} \, \frac{v}{v_0v_E}\, 
\Big\{ \, exp\Big[\, - \frac{(v-v_E)^2}{v_0^2 }\Big]\, - \,
exp\Big[\, - \frac{(v+v_E)^2}{v_0^2 }\Big]\, \Big\}
\label{4.15}  
\eeq
in agreement with Eq. (8.15) of Ref.~\cite{Jungm} In Eq. (\ref{4.11})
the nuclear parameters are implicit in the cross section $\sigma(v)$ 
given from Eq. (\ref{2.9}). The nuclear physics dependence of 
$\langle \Sigma\rangle $
could be disentangled by taking note of the extra velocity dependence 
of the coherent vector contribution in $\sigma(v)$
and introducing the parameters
\beq
\delta = \frac{2 v_E }{v_0}\, = \, 0.27,
\qquad  \psi = \frac{ v }{v_0}, \qquad  u = u_0 \psi ^2
\label{4.16}  
\eeq
where the quantity $u_0$ is the one entering the nuclear form factors of Eq. 
(\ref{2.16}) for $v=v_0$, which in this case is given by
\beq
u_0 = \frac{1}{2} \left( \frac{2\beta_0 m_x c^2}{(1+\eta)} 
\frac{b}{\hbar c}\right)^2,
\qquad \beta_0 = \frac{v_0}{c}
\label{4.17}  
\eeq
Afterwards, we can write Eq. (\ref{4.11}) as
\barr
\langle\Sigma\rangle&=&\Big(\frac{m_x}{m_N}\Big)^2\frac{\sigma_0}{(1+\eta)^2}
\nonumber  \\
&&\Big\{A^2 \Big[ \langle \beta^2 \rangle\Big(f^0_V-f^1_V \frac{A-2 Z}{A})^2 
\Big(J_0-\frac{2\eta+1}{2(1+\eta)^2}J_1\Big) 
\nonumber \\
&+& (f^0_S-f^1_S\frac{A-2 Z}{A})^2{\tilde J}_0\Big]  
\nonumber \\
&& + \Big( f^0_A \Omega_0(0)\Big)^2 J_{00}
+ 2 f^0_A f^1_A \Omega_0(0)\Omega_1(0) J_{01}
+ \Big( f^1_A \Omega_1(0)\Big)^2 J_{11} \Big\}
\label{4.18}  
\earr

If we assume that $J_{00}=J_{01}=J_{11}$, as seems to be the case for 
$^{207}Pb$, the spin dependent part of Eq. (\ref{4.14}) is reduced 
to the familiar expression 
$\Big[f^0_A \Omega_0(0) + f^1_A \Omega_1(0)\Big]^2 J_{11}$, where the quantity
in the bracket represents the spin matrix element at $q=0$.

The parameters ${\tilde J}_0$, $J_\rho$, $J_{\rho\sigma}$ describe the
scalar, vector and spin part of the counting rate, respectively,
and they are given by
\beq
{\tilde J}_0 (\lambda, u_0)  \,= \, \frac{2}{\sqrt{6\pi}}
\frac{e^{-\lambda^2}}{\lambda^2}
\int_0^{\infty}\psi e^{-\psi^2 } F_0 ( 2 \lambda \psi) I_0 (u_0\psi^2)d \psi
\label{4.19}  
\eeq
\beq
J_{\rho}(\lambda, u_0)  \,= \, \frac{2}{\sqrt{6\pi}}
\frac{e^{-\lambda^2}}{\lambda^2}
\int_0^{\infty}\psi^3 e^{-\psi^2 } F_0 ( 2 \lambda \psi) 
I_{\rho}(u_0\psi^2)d \psi
\label{4.20}  
\eeq
\beq
J_{\rho\sigma}(\lambda, u_0)  \,= \, \frac{2}{\sqrt{6\pi}}
\frac{e^{-\lambda^2}}{\lambda^2}
\int_0^{\infty}\psi e^{-\psi^2 } F_0 ( 2 \lambda \psi) 
I_{\rho\sigma}(u_0\psi^2)d \psi
\label{4.21}  
\eeq
\beq
\lambda \, = \, \frac{v_E}{v_0} = \Big[ \, 1 + \delta cos\alpha
sin\gamma + (\delta/2)^2 \Big]^{1/2}
\label{4.22}  
\eeq

The parameters $I_\rho$,  $I_{\rho\sigma}$ have been discussed in the 
previous section (see Figs. 3 and 4). The above integrals are functions of
 $\lambda$
and $u_0$. The latter depends on $v_0$, the nuclear parameters
and the LSP mass. These integrals can only be done numerically.
Since, however, $\lambda$ is close to unity, we can expand in powers of
$\delta$ and make explicit the dependence of these integrals on the earth's 
motion. Thus,
\beq
{\tilde J}_0(\lambda, u_0)  \,= \, \frac{2}{\sqrt{6\pi}}
B_1\,\Big[ \, {\tilde K}^{(0)}_0(u_0) 
+ \delta\, sin\gamma\, cos\alpha
 \, {\tilde K}^{(1)}_0(u_0)  \Big]
\label{4.23}  
\eeq
\beq
J_{\rho}(\lambda, u_0)  \,= \, \frac{2}{\sqrt{6\pi}}
B_2\,\Big[ \, K^{(0)}_{\rho}(u_0) 
+ \delta\, sin\gamma\, cos\alpha \, K^{(1)}_{\rho}(u_0)  \Big]
\label{4.24}  
\eeq
\beq
J_{\rho\sigma}(\lambda, u_0)  \,= \, \frac{2}{\sqrt{6\pi}}
B_1 \,\Big[ \,  K^{(0)}_{\rho\sigma}(u_0) 
+ \delta\, sin\gamma\, cos\alpha
 \,  K^{(1)}_{\rho\sigma}(u_0) \Big]
\label{4.25}  
\eeq

The integrals ${\tilde K}^{0}_{0}$, $K^{0}_{\rho}$ and  
$K^{0}_{\rho\sigma}$ are normalized so that they become unity at 
$u_0 =0$ (negligible momentum transfer). We find
\beq
B_1\, = \, \frac{1}{e}
\int_0^{\infty}\psi e^{-\psi^2 } F_0(2\psi) d \psi =
 \, \frac{1}{e} + 2\nu \, \approx \, 1.860
\label{4.26}  
\eeq
\beq
B_2\, = \, \frac{2}{3e}
\int_0^{\infty}\psi^3 e^{-\psi^2 } F_0(2\psi) d \psi
= \, \frac{2}{3} (\, \frac{3}{e} + 7 \nu ) \, \approx \, 4.220
\label{4.27}  
\eeq
with
\beq
\nu \, = \, \int_0^{1} e^{-t^2 } d t \,  \approx  \, 0.747
\label{4.28}  
\eeq
Furthermore,
\beq
{\tilde K}^{l}_0  \,= \, \frac{1}{eB_1}
\int_0^{\infty}\psi e^{-\psi^2 } F_l(2\psi) I_0 (u_0\psi^2)d \psi,
\qquad l=0,1
\label{4.30}  
\eeq
\beq
K^{l}_{\rho} = \frac{2}{3eB_2}\,
\int_0^{\infty}\psi^3 e^{-\psi^2 } F_l(2\psi) I_{\rho} (u_0\psi^2)d \psi,
\qquad l=0,1
\label{4.31}  
\eeq
\beq
K^{l}_{\rho\sigma} = \frac{1}{eB_1}\,
\int_0^{\infty}\psi e^{-\psi^2 } F_l(2\psi) I_{\rho\sigma} (u_0\psi^2)d \psi,
\qquad l=0,1
\label{4.32}  
\eeq
with $F_0(\chi)$ given in Eq. (\ref{4.12}) and
\beq
F_1(\chi) \, = \, 2\, \Big[ \,(\frac{\chi^2}{4} + 1) cosh\, \chi - 
\chi \,sinh \,\chi -1 \, \Big]
\label{4.33}
\eeq
The counting rate can thus be cast in the form
\beq
\Big<\frac{dN}{dt}\Big>=\Big<\frac{dN}{dt}\Big>_{0}(1+hcos\alpha)
\label{4.34}
\eeq
where $\big<\frac{dN}{dt}\big>_0$ is the rate obtained from the $l=0$ multipole 
and $h$ the amplitude of the oscillation, i.e. the ratio of the component
of the multipole $l=1$ to that of the multipole $l=0$. Below (see also Tables 
5(a),(b)) we compute separately the amplitude of oscillation for the scalar, 
vector and spin parts of the event rate i.e. the quantity
$h=\delta sin\gamma \, K^{1}(u_0)/K^{0}(u_0)$. 
Note the presence of the geometric factor $sin \gamma = 1/2$, which reduces 
the modulation effect.

In order to get some idea of the dependence of the counting rate on the 
earth's motion, we will evaluate the above expressions at $u_0=0$. We get
\beq
{\tilde K}^{0}_0  \,= \, 
K^{0}_{\rho\sigma} \approx K^{0}_{\rho}   \,= \, 1
\label{4.35}
\eeq
\beq
{\tilde K}^{1}_0  \,= \, 
K^{1}_{\rho\sigma} \,= \, \frac{\nu}{1/e +2\nu} \approx 0.402
\label{4.36}
\eeq
\beq
K^{1}_{\rho}   
 \,= \, \frac{3/(2e) + (11/2)\nu/2}{3/e +7\nu} \approx 0.736
\label{4.37}
\eeq
Thus, for $sin\gamma \approx 0.5$
\beq
{\tilde J}_0 \approx J_{\rho\sigma} = \frac{2}{\sqrt{6\pi}}\, 1.860 
(1 + 0.054 cos\alpha) = 0.857(1+ 0.054 cos\alpha)
\label{4.38}
\eeq
\beq
J_{\rho} = \frac{2}{\sqrt{6\pi}}\, 4.220 (1 + 0.099 cos\alpha)
= 1.944 (1+ 0.099 cos\alpha)
\label{4.39}
\eeq

We see that, the modulation of the detection rate due to the earth's motion
is quite small ($h \approx 0.05$). The corresponding amplitude of 
oscillation in the coherent vector contribution, Eq. (\ref{4.38}), is a bit 
bigger ($h \approx 0.10$). However, this contribution is suppressed due to 
the Majorana nature of LSP (through the factor $\beta^2$). The modulation due 
to the Earth's rotation is expected to be even smaller.

The exact $K^{l}$ integrals, for the l=0 and l=1, are shown in Figs.
 5(a),(b), (c). The most important of these integrals, those of 
Eq. (\ref{4.30}) associated with the
scalar interaction, are shown in Fig. 5(a). In Fig. 5(b)  we present the 
integrals of Eq. (\ref{4.31}) for $\rho=0 $ associated with the vector 
interaction (the integral for  
$\rho=1 $ is analogous but it is less important). Finally in Fig. 5(c) the    
integrals of Eq. (\ref{4.32}) for $\rho=1 $ and $ \sigma=1 $ are shown.
The others are practically indistinguishable from these and are not shown
(see Ref.~\cite{KVcdm}). 

Before closing this section we should mention that, the folding procedure can
also be applied in the differential rate in order to obtain the corresponding 
convoluted expression for $d\sigma/d\Omega$, i.e. before doing the angular 
integration in Eq. (\ref{2.6}) and obtain the total cross section Eq. 
(\ref{2.9}). In fact this may be important since the modulation effect in
the total cross section is small due to cancellations. In fact preliminary
results indicate that the modulation effect can get as high as $20 \%$.
The high value occurs, unfortunately, in the regions where the total
differential rate becomes too small.

\section{Results and discussion }

The three basic ingredients of our calculation were 
the input SUSY parameters, a quark model for the nucleon and the structure
of the nuclei involved.
The input SUSY parameters used for the results presented in Tables 1 and 2
have been calculated in a phenomenologically allowed 
parameter space (cases \#1, \#2, \#3 of Ref.~\cite{Kane} and cases \#4-9
of Ref.~\cite{Casta}

For the coherent part (scalar and vector) we used realistic nuclear
form factors and studied three nuclei, representatives
of the light, medium and heavy nuclear isotopes ($Ca$, $Ge$ and $Pb$).
In Tables 5(a),(b) and 6 we show the results obtained for three different quark
models denoted by A (only quarks u and d) and B, C (heavy quarks
in the nucleon).
We see that the results vary substantially and are very sensitive to 
the presence of quarks other than u and d into the nucleon.

The spin contribution, arising from the axial current,
was computed in the case of $^{207}Pb$ system. 
For the isovector axial coupling the transition from the quark to
the nucleon level is trivial (a factor of $g_A=1.25)$. For the
isoscalar axial current we considered two possibilities 
depending on the portion of the nucleon spin which is attributed to the
quarks, indicated by EMC and NQM.~\cite{KVprd}
The ground state wave function of $^{208}Pb$ was obtained
by diagonalizing the nuclear Hamiltonian~\cite{KV90}-\cite{Kuo} 
in a 2h-1p space which is standard for this doubly magic nucleus.   
The momentum dependence of the matrix elements was taken into account and all 
relevant multipoles were retained (here only $\lambda=0$ and $\lambda=2$).

In  Table 4(a), we compare the spin matrix elements at $q=0$ for the 
most popular targets considered for LSP detection $^{207}Pb$, 
$^{73}Ge$ and $^{29}Si$. We see that, even though the spin matrix elements 
$\Omega^2$ are even a factor of three smaller than those for $^{73}Ge$ obtained 
in Ref.~\cite{Ress} (see Table 4(a)), their contribution to the total
cross section is almost the same (see Table 4(b)) for LSP masses around 
$100 \, GeV$. Our final results for the quark models (A, B, C, NQM, EMC)
are presented in Tables 5(a),(b) for SUSY models \#1-\#3~\cite{Kane}
and Table 6 for SUSY models \#4-\#9.~\cite{Casta}

\section{Conclusions }

In the present study we found that for heavy LSP and heavy nuclei
the results are sensitive to the momentum transfer as well as to the
LSP mass and other SUSY parameters. From the Tables 5(a),(b) and 6  we see that,
the results are also sensitive to the quark structure of the nucleon.
We can, however, draw the following general conclusions.

(i) The coherent scalar (associated with Higgs exchange) for model 
A (u and d quarks only) is comparable to the vector coherent
contribution. Both are at present undetectable. For models B and C
(heavy quarks in the nucleon) the coherent scalar contribution is
dominant. Detectable rates $\langle dN/dt\rangle_0\ge 100\,\,y^{-1}Kg^{-1}$ 
are possible in a number of models with light LSP.

(ii) The folding of the total event rate with the velocity distribution
provides the total modulation effect $h$. In all cases it is small, 
less than $\pm 5\%$.

(iii) The spin contribution is sensitive to the nuclear structure.
It is undetectable if the LSP is primarily a gaugino. 

\bigskip
\bigskip
{\it Acknowledgements:} One of the authors (JDV) would like to
acknowledge partial support of this work by $\Pi N E \Delta$ 1895/95 
of the Greek secretariat for Research, by TMR No ERB FMAX-CT96-0090
of the European Union, the INTAS 93-1648 program for travel support to
attend the Workshop and the workshop organizers for hospitality.

\bigskip


\newpage
\noindent
{\bf Table  1.} The essential parameters describing the LSP and Higgs.
For the definitions see the text.

\vspace{0.2cm}
\begin{center}
\begin{tabular}{|l|lllllllll|}
\hline
\hline
 & & & & & & & & &  \\
 Solution & $\#$1 & $\#$2 & $\#$3 & $\#$4 & $\#$5 & $\#$6 & $\#$7 & $\#$8 
 & $\#$9 \\
\hline
 & & & & & & & & &  \\
$m_x \,(GeV)$ & 126 & 27 & 102 & 80 & 124 & 58 & 34 & 35 & 50 \\
$m_h$ & 116.0& 110.2& 113.2& 124.0& 121.0& 105.0& 103.0 & 92.0& 111.0 \\
$m_H$ & 345.6& 327.0& 326.6& 595.0& 567.0& 501.0& 184.0& 228.0& 234.0 \\
$m_A$ & 345.0& 305.0& 324.0& 594.0& 563.0& 497.0& 179.0& 207.0& 230.0 \\
tan2$\alpha$& 0.245& 6.265& 0.525& 0.410& 0.929& 0.935& 0.843& 1.549& 0.612 \\
tan$\beta$& 10.0& 1.5&  5.0&   5.4&   2.7&   2.7&   5.2&   2.6&   5.3 \\
\hline
\hline
\end{tabular}
\end{center}

\vspace{0.8cm}

\noindent
{\bf Table 2(a).} The coupling constants entering ${\cal L}_{eff}$, eqs.
(\ref{eq:eg 41}), (\ref{eq:eg.42}) and (\ref{eq:eg.45}) of the text, for 
solutions \#1 - \#3.

\vspace{0.2cm}
\begin{center}
\begin{tabular}{|l|ccc|}
\hline
\hline
 &  &  &  \\
Quantity & Solution $\#1$ & Solution $\#2$ & Solution $\#3$  \\
\hline
 &  &  &  \\
$\beta f^0_V$ &$1.746\times10^{-5}$&$2.617\times10^{-5}$
 & $2.864\times10^{-5}$ \\
$f^1_V/f^0_V$ &-0.153 & -0.113 &-0.251 \\
$f^0_S(H)$ (model A) &$1.31\times10^{-5}$&$1.30\times10^{-4}$ 
   &$1.38\times10^{-5}$ \\
$f^1_S/f^0_S$ (model A) &-0.275 &-0.107 &-0.246 \\
$f^0_S(H)$ (model B) &$5.29\times10^{-4}$&$7.84\times10^{-3}$ 
   &$6.28\times10^{-4}$ \\
$f^0_S(H)$ (model C) &$7.57\times10^{-4}$&$7.44\times10^{-3}$ 
   &$7.94\times10^{-4}$ \\
$f^0_A (NQM) $& $0.510\times10^{-2}$&$3.55\times10^{-2}$
 & $0.704\times10^{-2}$ \\
$f^0_A (EMC) $& $0.612\times10^{-3}$&$0.426\times10^{-2}$
 & $0.844\times10^{-3}$ \\
$f^1_A $& $1.55\times10^{-2}$&$5.31\times10^{-2}$
 & $3.00\times10^{-2}$ \\ 
\hline
\hline
\end{tabular}
\end{center}

\vspace{0.5cm}

\noindent
{\bf Table  2(b).} The same as in Table 1(a) for solutions \#4 - \#9.

\vspace{0.2cm}
\begin{center}
\begin{tabular}{|l|llllll|}
\hline
\hline
  & & & & & &  \\
 Solution   &  $\#$4 & $\#$5  & $\#$6 & $\#$7  & $\#$8  & $\#$9 \\
\hline
  & & & & & &  \\
$\langle \beta^2\rangle^{1/2} f_V^0 $& 0.225 $10^{-4}$ & 0.190 $10^{-4}$&
0.358 $10^{-4}$ & 0.108 $10^{-4}$ & 0.694 $10^{-4}$ & 0.864 $10^{-4}$  \\
$f_V^1/f^0_V$ & -0.0809 & -0.0050 & -0.0320 &
                -0.0538 & -0.0464 & -0.0369   \\
$f_S^0(A)$  &-0.179 $ 10^{-4}$&-0.236 $ 10^{-4}$&
             -0.453 $ 10^{-4}$&-0.266 $ 10^{-4}$&
             -0.210 $ 10^{-3}$&-0.131 $ 10^{-3}$  \\
$f_S^0(B)$  &-0.531 $ 10^{-2}$&-0.145 $ 10^{-2}$&
             -0.281 $ 10^{-2}$&-0.132 $ 10^{-1}$&
             -0.117 $ 10^{-1}$&-0.490 $ 10^{-2}$  \\
$f_S^0(C)$  &-0.315 $ 10^{-2}$&-0.134 $ 10^{-2}$&
             -0.261 $ 10^{-2}$&-0.153 $ 10^{-1}$&
             -0.118 $ 10^{-1}$&-0.159 $ 10^{-2}$  \\
$f_S^1(A)$  &-0.207 $ 10^{-5}$&-0.407 $ 10^{-5}$&
              0.116 $ 10^{-4}$& 0.550 $ 10^{-4}$&
              0.307 $ 10^{-4}$& 0.365 $ 10^{-4}$  \\
$f_A^0(NQM)$& 6.950 $ 10^{-3}$& 5.800 $ 10^{-3}$&
              1.220 $ 10^{-2}$& 3.760 $ 10^{-2}$&
              3.410 $ 10^{-2}$& 2.360 $ 10^{-2}$  \\
$f_A^0(EMC)$& 0.834 $ 10^{-3}$& 0.696 $ 10^{-3}$&
              0.146 $ 10^{-2}$& 0.451 $ 10^{-2}$&
              0.409 $ 10^{-2}$& 0.283 $ 10^{-2}$  \\
$f_A^1$     & 2.490 $ 10^{-2}$& 1.700 $ 10^{-2}$&
              3.440 $ 10^{-2}$& 2.790 $ 10^{-1}$&
              1.800 $ 10^{-1}$& 2.100 $ 10^{-1}$  \\
\hline
\hline
\end{tabular}
\end{center}

\newpage
\noindent
{\bf Table 3.} The quantity $q_0$ (forward momentum transfer) in units of
$fm^{-1}$ for three values of $m_1$ and three typical nuclei.
In determining $q_0$ the value $\langle \beta^2 \rangle^{1/2} =10^{-3}$ was
employed.

\vspace{0.2cm}
\begin{tabular}{|c|ccc|}
\hline
\hline
  &  &  &  \\
& \multicolumn{3}{|c|}{ $q_0$ ($fm^{-1}$)}  \\
\hline
  &  &  &  \\
Nucleus & $m_1=30.0 \, GeV$ & $m_1=100.0 \, GeV$ & $m_1=150.0 \, GeV$ \\
\hline
  &  &  &  \\
  $Ca$  &  0.174   &  0.290 &  0.321 \\
  $Ge$  &  0.215   &  0.425 &  0.494 \\
  $Pb$  &  0.267   &  0.685 &  0.885 \\ 
\hline
\hline
\end{tabular}

\vspace{0.8cm}

\noindent
{\bf Table 4(a).} Comparison of the static spin matrix elements for
three typical nuclei, $Pb$ (present  calculation) and $^{73}Ge$, $^{29}Si $
(see Ref. \cite{Ress}).

\vspace{0.2cm}
\begin{center}
\begin{tabular}{|l|rrr|}
\hline
\hline
  &  &  &  \\
Component & $^{207}Pb_{{1/2}^-}$ & $^{73}Ge_{{9/2}^+}$ 
 & $^{29}Si_{{1/2}^+}$ \\
\hline
  &  &  &  \\
$\Omega^2_1(0)$ & 0.231 & 1.005 & 0.204 \\
$\Omega_1(0) \Omega_0(0)$ & -0.266 & -1.078 & -0.202 \\
$\Omega^2_0(0)$ & 0.305 & 1.157 & 0.201 \\
\hline
\hline
\end{tabular}
\end{center}

\vspace{0.5cm}

\noindent
{\bf Table 4(b).}
Ratio of spin contribution ($^{207}Pb/^{73}Ge$) at the relevant
momentum transfer with the kinematical factor $1/(1+\eta)^2, \,\, 
\eta=m/A m_N.$

\vspace{0.2cm}
\begin{center}
\begin{tabular}{|c|lllllllll|}
\hline
\hline
 & & & & & & & &  &   \\
 Solution & $\#$1 & $\#$2 & $\#$3 & $\#$4 & $\#$5 & $\#$6 & $\#$7 & $\#$8 
 & $\#$9 \\
\hline
 & & & & & & & &  &   \\
$m_x \,(GeV)$ & 126 & 27 & 102 & 80 & 124 & 58 & 34 & 35 & 50 \\
\hline
 & & & & & & & &  &   \\
NQM & 0.834 & 0.335 & 0.589 & 0.394 & 0.537 & 0.365 & 0.346 & 0.337 & 0.417 \\
EMC & 0.645 & 0.345 & 0.602 & 0.499 & 0.602 & 0.263 & 0.341 & 0.383 & 0.479 \\
\hline
\hline
\end{tabular}
\end{center}

\newpage
\noindent
{\bf Table 5(a).} The quantity $\langle dN/dt\rangle_0$ in $y^{-1}Kg^{-1}$
and the modulation  parameter h for the coherent vector and scalar 
contributions in the  cases \#1 - \#3 and for three typical nuclei.

\vspace{0.2cm}
\begin{center}
\begin{tabular}{|rl|cc|lrrc|}
\hline
\hline
& & & & & & &   \\
 & & \multicolumn{2}{|c|}{Vector $\,\,\,$ Contribution}  &
     \multicolumn{4}{|c|}{Scalar $\,\,\,$ Contribution}  \\
\hline 
& & & & & & &   \\
& & $\langle dN/dt\rangle_0$ &$h$ &\multicolumn{3}{c}{$\langle 
dN/dt\rangle_0 $}& $h$ \\ 
\hline
& & & & & & &   \\
& \multicolumn{1}{c|}{Case} &$(\times 10^{-3})$ &  &
    \multicolumn{1}{c}{ Model $\,\,$ A }& 
    \multicolumn{1}{c}{ Model $\,\,$ B }& 
    \multicolumn{1}{c}{ Model $\,\,$ C }&  \\ 
\hline
& & & & & & &   \\
  &$\#1$& 0.264 &0.029 &$0.151\times 10^{-3}$ &  0.220 &  0.450 &-0.002 \\
Pb&$\#2$& 0.162 &0.039 &$0.410\times 10^{-1}$ &142.860 &128.660 & 0.026 \\
  &$\#3$& 0.895 &0.038 &$0.200\times 10^{-3}$ &  0.377 &  0.602 &-0.001 \\
\hline
& & & & & & &   \\
  &$\#1$& 0.151 &0.043 &$0.779\times 10^{-4}$ &  0.120 &  0.245 & 0.017 \\
Ge&$\#2$& 0.053 &0.057 &$0.146\times 10^{-1}$ & 51.724 & 46.580 & 0.041 \\
  &$\#3$& 0.481 &0.045 &$0.101\times 10^{-3}$ &  0.198 &  0.316 & 0.020 \\
\hline
& & & & & & &   \\
  &$\#1$& 0.079 &0.053 &$0.340\times 10^{-4}$ &  0.055 &  0.114 & 0.037 \\
Ca&$\#2$& 0.264 &0.060 &$0.612\times 10^{-2}$ & 22.271 & 20.056 & 0.048 \\
  &$\#3$& 0.241 &0.053 &$0.435\times 10^{-4}$ &  0.090 &  0.144 & 0.038 \\
\hline
\hline
\end{tabular}
\end{center}

\vspace{0.5cm}

\noindent
{\bf Table 5(b).} The spin contribution in the $LSP-^{207}Pb$ scattering 
for two cases: EMC data and NQM Model for solutions $\#1, \#2, \#3$. 

\vspace{0.2cm}
\begin{center}
\begin{tabular}{|l|ll|lc|}
\hline
\hline
& & & &   \\
& \multicolumn{2}{|c}{\hspace{1.2cm}EMC \hspace{.2cm} DATA} \hspace{.8cm} &
 \multicolumn{2}{|c|}{\hspace{1.2cm}NQM \hspace{.2cm} MODEL} \\ 
\hline
& & & &   \\
Solution & \hspace{.2cm}$\langle dN/dt \rangle _0$ $(y^{-1} Kg^{-1})$
  \hspace{.2cm} & $ h $ & 
$\hspace{.2cm} \langle dN/dt \rangle_0 $ $ (y^{-1} Kg^{-1})$ & $ h $ \\ 
\hline
& & & &   \\
$\#1  $  &$0.285\times 10^{-2}$& 0.014 &$0.137\times 10^{-2}$& 0.015  \\
$\#2  $  & 0.041               & 0.046 &$0.384\times 10^{-2}$& 0.056  \\
$\#3  $  & 0.012               & 0.016 &$0.764\times 10^{-2}$& 0.017  \\
\hline
\hline
\end{tabular}
\end{center}

\newpage
\noindent
{\bf Table 6.} The same parameters as in Tables 5(a) for $Pb$ for the solutions 
 $\#4-\#9$. Cases $\#8, \#9$ are no-scale models. The values of 
$\langle dN/dt \rangle_0$ for Model A and the Vector part must be multiplied by
$\times 10^{-2}$.

\vspace{0.2cm}
\begin{center}
\begin{tabular}{|l|crrr|lr|lll|}
\hline
\hline
 & & & & & & & & &   \\
  & \multicolumn{4}{|c|}{Scalar$\,$ Part} & 
    \multicolumn{2}{c|}{Vector $\,$ Part} &
    \multicolumn{3}{c|}{Spin   $\,$ Part} \\
\hline 
 & & & & & & & & &   \\
& \multicolumn{3}{c}{$\big<\frac{dN}{dt}\big>_0$}& $h$ & 
$\big<\frac{dN}{dt}\big>_0$ & $h$ & \multicolumn{2}{c}
{$\big<\frac{dN}{dt}\big>_0$} & $h$ \\ 
\hline
 & & & & & & & & &   \\
Case & A & B & C & & & & EMC & NQM & \\ 
\hline
 & & & & & & & & &   \\
$\#4 $& 0.03 & 22.9& 8.5 & 0.003& 0.04 & 0.054
 & $0.80\, 10^{-3}$& $0.16\, 10^{-2}$& 0.015 \\
$\#5 $& 0.46 & 1.8& 1.4 & -0.003& 0.03 & 0.053
 & $0.37\, 10^{-3}$& $0.91\, 10^{-3}$& 0.014 \\
$\#6 $& 0.16 & 5.7& 4.8 & 0.007& 0.11 & 0.057
 & $0.44\, 10^{-3}$ & $0.11\, 10^{-2}$& 0.033 \\
$\#7 $&  4.30 & 110.0& 135.0 & 0.020& 0.94 & 0.065
 & 0.67 & 0.87 & 0.055 \\
$\#8 $& 2.90 & 73.1& 79.8 & 0.020& 0.40 & 0.065
 & 0.22 & 0.35 & 0.055 \\
$\#9 $&  2.90 & 1.6& 1.7 & 0.009& 0.95 & 0.059
 & 0.29 & 0.37 & 0.035 \\
\hline
\hline
\end{tabular}
\end{center}

\newpage

\vspace*{0.5cm}

\centerline{\large \bf FIGURE CAPTIONS}

\bigskip
\bigskip

\noindent
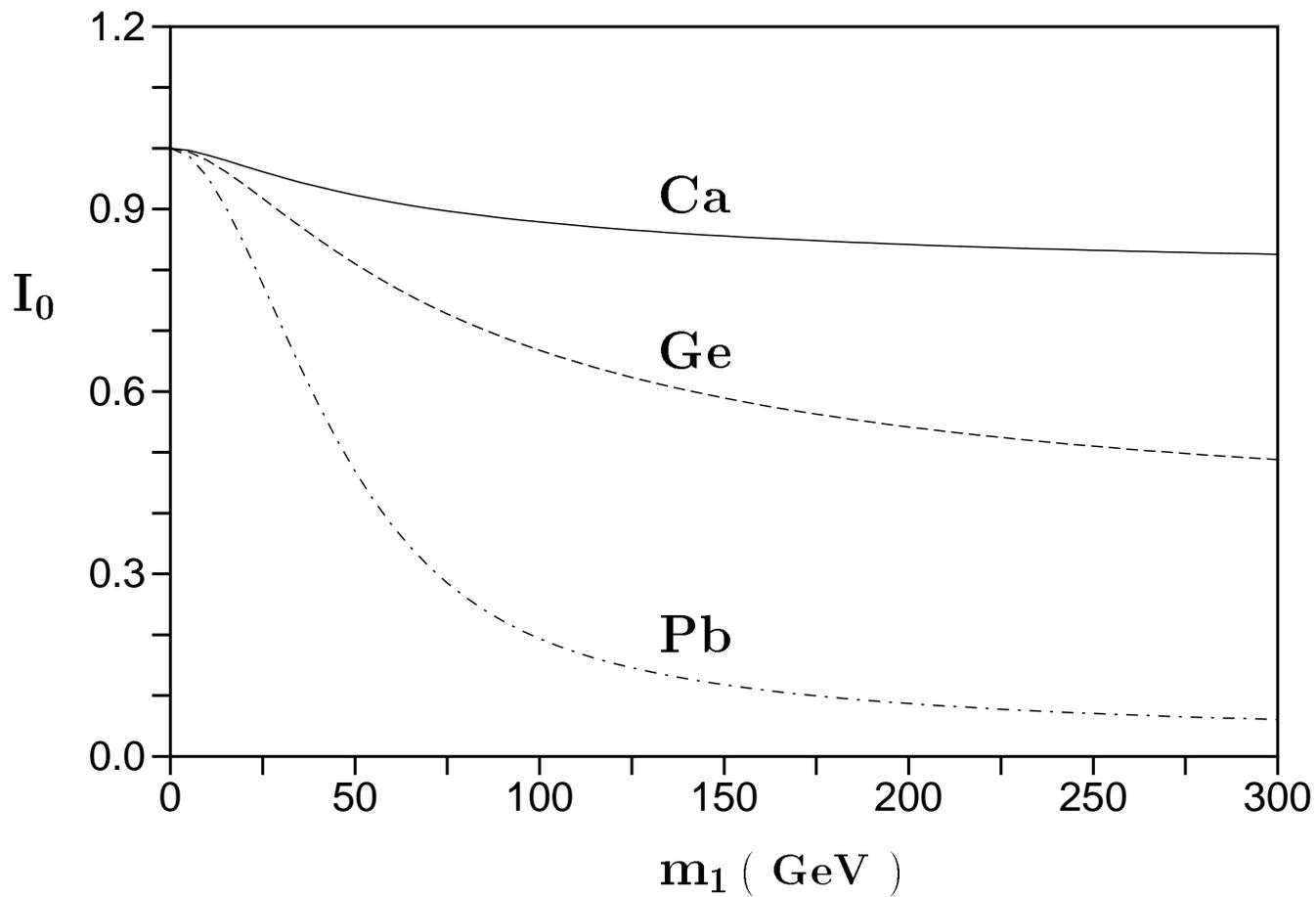
{\bf Fig. 1.}
\medskip
Two diagrams which contribute to the elastic scattering of LSP with nuclei:
Z-exchange in Fig. 1(a) and s-quark exchange in Fig. 1(b).  Due to the Majorana
nature of LSP only its pseudovector coupling contributes. $J_\lambda$ can be
parametrized in terms of  four form factors $f^0_V, f^1_V, f^0_A, f^1_A$. The
scalar terms arising from s-quark mixing are negligible in the SUSY parameter 
space considered here.

\bigskip
\bigskip

\noindent
{\bf Fig. 2.}
\medskip
The same as in Fig. 1, except that the intermediate Higgs exchange is
considered. This leads to an effective scalar interaction with two form
factors $f^0_S$, (isoscalar) and  $f^1_S$ (isovector).

\bigskip
\bigskip

\noindent
{\bf Fig. 3.}
\medskip
The integrals $I_0$ which describe the dominant scalar contribution 
(coherent part) of the total cross section as a function of the LSP mass 
($m_x \equiv m_1$), for three typical nuclei: $Ca$, $Ge$ and $Pb$.
The value $\langle \beta^2 \rangle^{1/2} = 10^{-3}$ was used.

\bigskip
\bigskip

\noindent
{\bf Fig. 4.}
\medskip
(a) Plot of the integrals $I_{11}$ as a function of the LSP mass $m_x \equiv 
m_1$. This integral gives the spin contribution to the LSP-nucleus total 
cross section for $^{207}Pb$. The integrals $I_{00}$ and $I_{01}$ are similar.
(b) Plot of the integrals $I_{11}(u)$ and $I_0(u)$ for $Pb$. Note that $I_{11}$
is quite a bit less retarded compared to $I_{0}$.
For definitions see the text.

\bigskip
\bigskip

\noindent
{\bf Fig. 5.}
\medskip
Contributions of K integrals (for l=0 and l=1) entering the event rate due to
earth's revolution around the sun: ${\tilde K}^l_0$ in Fig. 5(a), $K^l_0$ in
Fig. 5(b) and $K^l_{11}$ in Fig. 5(c). The other integrals $K^l_{00}$ and
$K^l_{01}$ are similar to $K^l_{11}$. 

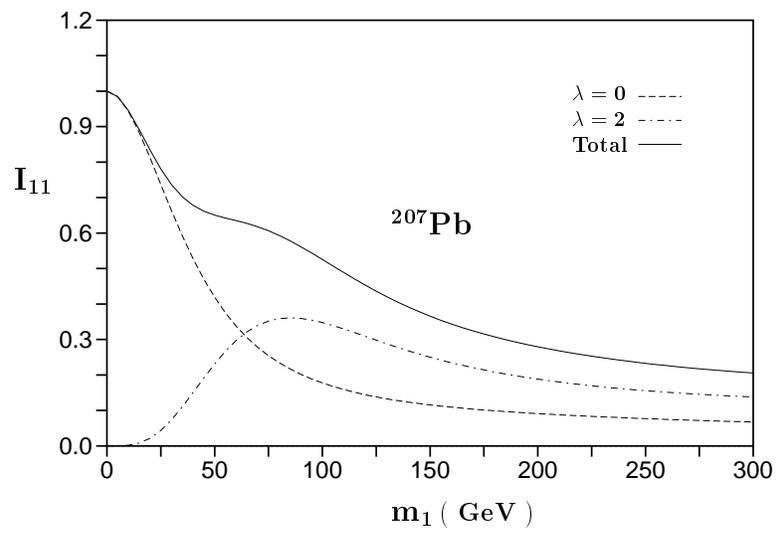
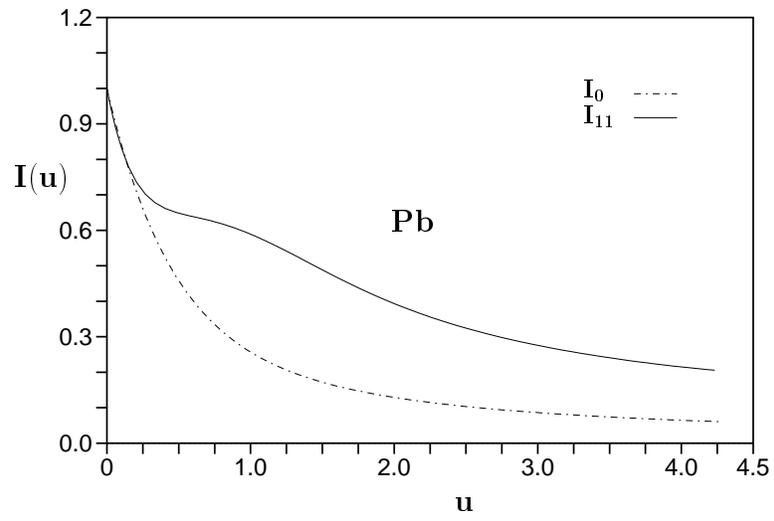
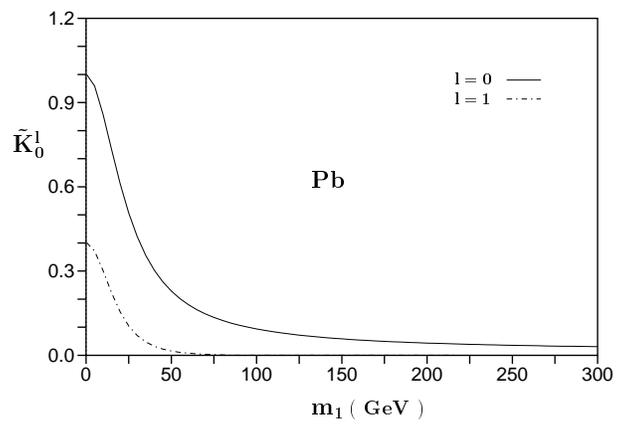
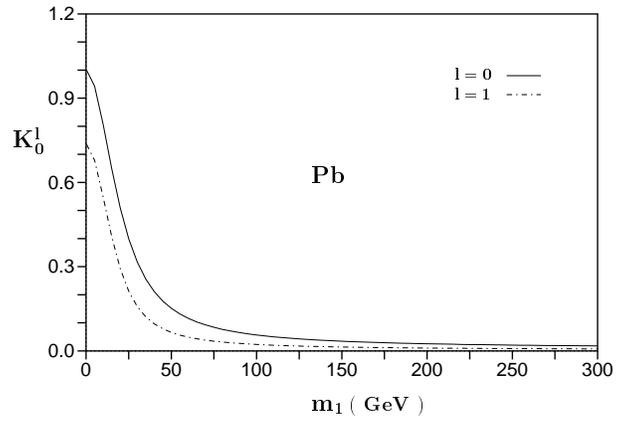
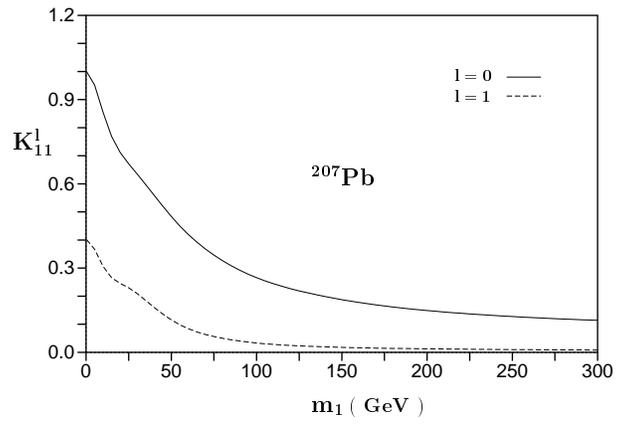
\end{document}